\documentclass{llncs}
\usepackage{bm,amsmath,amssymb}
\usepackage{amsfonts}
\usepackage{fancyhdr}
\usepackage{mathrsfs}
\usepackage{graphicx}
\usepackage{epstopdf}
\usepackage{color}
\usepackage{latexsym}
\usepackage{booktabs}
\usepackage{multirow}
\usepackage{subfigure}
\usepackage[bookmarks = false]{hyperref}
\usepackage{url}

\usepackage{algorithm}
\usepackage{algorithmicx}
\usepackage{algpseudocode}

\usepackage{array}

\begin{document}
\title{Notes on Rainbow Distinguished Point Method}
\author{Wenhao Wang}
\institute{State Key Laboratory Of Information Security,\\
\and Institute of Information Engineering, CAS, Beijing, China.\\
\email{wangwenhao@iie.ac.cn}}

\maketitle
\begin{abstract}
This paper clarifies the method described in our previous paper (DOI: 978-3-319-02726-5\_21), namely rainbow distinguished point method, and give revised theoretical and experimental results which shows rainbow distinguished point method behaves inferiorly to other time memory tradeoff methods.
\end{abstract}

\section{Introduction}
%结合密码破解，如windows密码破解等

Time memory tradeoff algorithm is introduced by Hellman in 1980s~\cite{hellman}. It is a generic method to invert one-way functions and is useful in symmetric cipher design and analysis. After that, a lot of work has been done to improve the efficiency of time memory tradeoff algorithms, such as perfect tables and distinguished point method. Rainbow tradeoff, introduced by Oechslin at CRYPTO 2003, is the most widely known method, which applies to crack various password based systems~\cite{rainbow}.

Distinguished point method is put forward by Rivest to reduce the time required to access pre-computed tables. Our previous paper (\cite{wang2011improvement}) presented a combination of DP method and rainbow tradeoff, called rainbow distinguished point method. It used a rather rough approximation of $\widetilde{m}_i$. It shows that rainbow distinguished point method is NOT recommended compared to other TMTO algorithms.

This paper give a full description of the offline and online phase of rainbow distinguished point method. With a more accurate analysis, we show how to find optimal parameters and that rainbow distinguished point method behaves inferiorly to other time memory tradeoff methods.
As a matter of fact, recent works on this subject~\cite{kim2013analysis,haghighi2014practical} have shown that the rainbow fuzzy tradeoff (another combination of DP method and rainbow tradeoff) performs the best among all existing tradeoff algorithms.

The rest of this paper are as follows. We first give a full description of the offline phase and online phase of rainbow distinguished point method. Then we theoretically analyze the tradeoff curve and show how to choose optimal parameters. In the last, we present experimental results and make a comparison with some of the existing tradeoff algorithms.

\section{Description of Rainbow Distinguished Point Method}
The procedure of rainbow distinguished point method consists of the offline phase and online phase.

\subsection{Offline phase}
Fix a distinguished property, and the probability of a random selected $x \in \mathcal{N}$ to be a distinguished point is $1/t$. We refer to $t$ as the chain length. In general, the distinguished points set
$$
C = \{x ~\vert~ x \in \mathcal{N} \mbox{~and~MSB}_{k}(x) = {\bf{0}}\},
$$
where $\mbox{MSB}_k(x)$ denote the first $k$ bits of $x$, i.e. $t = 2^k$. A chain length bound ~$\hat{t} = ct$ is set to discard any chain exceeds a maximum chain length.
During the offline phase, $l$ pre-computed tables consisting of $m$ pre-computed chains are constructed. Every column of all pre-computed tables uses a different reduction function. We denote the $s$-th column of the $i$-th table by $r_{i,s}~(~0 \le i \le l-1,~1 \le s < \hat{t}~)$ and $r_{i,s}(y)$ is defined as
$$
r_{i,s}(y) := (y + i\hat{t} + s)\mbox{ mod }N \mbox.
$$
We take ~$f_{i,s} = r_{i,s} \circ f$ for short. For the $i$-th table,

\noindent\fbox{
\parbox{\textwidth}{
\begin{itemize}
\item[1.~]
Randomly select ~$\widetilde{m}_0$~different {\bf{starting points}}~$\mbox{SP}_{i,1},~\mbox{SP}_{i,2},~\cdots,~\mbox{SP}_{i,\widetilde{m}_0}$.
\item[2.~]
Construct pre-computed chains from every starting points. For example，the~$j~(~1 \le j \le \widetilde{m}_0~)$~chain ~$X^i_{j,1} = \mbox{SP}_{i,j}$; For~$s = 1,~2,~\cdots$，compute ~$Y^i_{j,s} = f(X^i_{j,s})$~ and ~$X^i_{j,s+1} = r_{i,s}(Y^i_{j,s}) = f_{i,s}(X^i_{j,s})$ iteratively. The pre-computed chain continue to compute until some distinguished point appears. Denote the {\bf{ending points}} by ~$\mbox{EP}_{i,j}$. If the pre-computed chain exceeds the maximum length ~$\hat{t}$~ before arriving at some distinguished point, then we discard this pre-computed chain and continue to compute the next chain.
\item[3.~]
We construct ~$m_0$~ pre-computed chains, which comprises a pre-computed ‘’matrix’’.
\item[4.~]
The tuple of starting points, ending points and the corresponding chain length (~$\mbox{SP}_{i,j},len_{i,j},\mbox{EP}_{i,j}$~)~$(~1 \le j \le m_0~)$ of ~$m_0$~ chains are sorted according the chain length (or the ending points) and stored as the $i$-th pre-computed table.
\end{itemize}
}}

\subsection{Online phase}
During the online phase, the $l$ pre-computed tables are searched in parallel (as depicted in Alg.~\ref{algo:rainbowdp} ).

\begin{algorithm}[h!]
%\fbox{
%\parbox{\textwidth}{
\caption{online searching algorithm of rainbow distinguished point method}
\begin{algorithmic}[1]
\Require
$y^* \in \mathcal{N}$；
The expression or description of ~$f$~;

Number of pre-computed tables, number of chains in each table, chain length~$t$~, chain length bound~$\hat{t}$;

Distinguished points set~$C$;

Pre-computed tables: $(\mbox{SP}_{i,j}, len_{i,j}, \mbox{EP}_{i,j})_{1 \le i \le l, 1 \le j \le m}$~;
\Ensure
Success: ~$x^*$ satisfying $f(x^*) = y^*$. Or failure.
\For {$s = 1$ to $\hat{t}$}
\For {$i = 1$ to $l$}
\State $q \gets r_{i,\hat{t}-s+1}(y^*)$；
\For {$k = s$ to $1$}
\State $q \gets f_{i,\hat{t}-k}(q)$；
\If {~$q \in C$}
\State Search $q$ among the ending points of chains with length $(~\hat{t}-k+1~)$ in the $i$-th table;
\For {$j \in \{j~\vert~len(i,j) \mbox{equals} (~\hat{t}-k+1~) \mbox{and} q\mbox{equals}~\mbox{EP}_{i,j}\}$}
\State {$x = \mbox{SP}_{i,j}$};
\For {$u = 1$ to $\hat{t}-s$}
\State {$x \gets f_{i,u}(x)$};
\EndFor
\If {~$f(x)$~equals~$y^*$~}
\State $x^* \gets x$;
\State \Return Success;
\EndIf
\EndFor
\EndIf
\EndFor
\EndFor
\EndFor
\State \Return Failure;
\end{algorithmic}
%}}
\label{algo:rainbowdp}
\end{algorithm}

A pre-image of $y^*$ can be found during the online phase only if $y^*$ appears in the pre-computed matrix. The probability of the online phase is called the success probability of rainbow distinguished point method. An execution of the outer loop $s$ is called the $s$-th (online) iterating search. If $q$ appears as a ending point of a pre-computed table, we say an ‘’alarm’’ happens. Furthermore if $q$ appears as a ending point of a pre-computed table, and the pre-image is NOT found after regenerating the pre-computed chain, way say a ‘’false alarm’’ happens.

\section{Optimizing the Tradeoff Parameters}

\subsection{Theoretical Analysis}
Denote by $\widetilde{m}_0$ the number of random selected starting points. Denote by $\widetilde{m}_i$ (or $m_i$) the number of different elements in the $i$-th column of the pre-computed matrix, with $\widetilde{m}_0$ (or $m_0$) lines, before (or after) removing the chains not arriving at distinguished points. Denote by $D_{pc} = m_0tl/N$ the pre-computation coefficient.

\begin{lemma}
Let $H = 2/(2+\frac{\widetilde{m}_0t}{N})$. Then
$$
\begin{aligned}
\widetilde{m}_i = \frac{\widetilde{m}_0H}{e^{\frac{i}{t}}-(1-H)} , \\
{m}_i = \widetilde{m}_i \cdot (1-e^{i/t-c}) \mbox,
\end{aligned}
$$
where $c = \hat{t}/t$.
\end{lemma}

\proof
The probability of the elements in the $i$-th column to be a distinguished point is $\frac{1}{t}$. Thus
$$
\begin{aligned}
\widetilde{m}_{i+1}&=N(1-(1-\frac 1N)^{\widetilde{m}_i(1-\frac 1t)}) \\
&=N\bigg(\frac{\widetilde{m}_i(1-\frac{1}{t})}{N} - \frac{\widetilde{m}^2_i(1-\frac{1}{t})^2}{2N^2} + \mathcal{O}\big(\frac{\widetilde{m}^3_i(1-\frac{1}{t})^3}{N^3}\big)\bigg) \\
& \approx \widetilde{m}_i - \frac{\widetilde{m}_i}{t}-\frac{\widetilde{m}^2_i}{2N} \mbox,
\end{aligned}
$$
Discard the terms of order less than $\widetilde{m}_i/t$, then it turns out that
$$
\widetilde{m}_{i+1} - \widetilde{m}_{i} = - \frac{\widetilde{m}_i}{t}-\frac{\widetilde{m}^2_i}{2N} \mbox.
$$
Solve the differential equation
$$
\frac{d~{\widetilde{m}_i}}{d~i} = - \frac{\widetilde{m}_i}{t}-\frac{\widetilde{m}^2_i}{2N}\mbox,
$$
and let $H = 2/(2+\frac{\widetilde{m}_0t}{N})$, then we have
$$
\widetilde{m}_i = \frac{\widetilde{m}_0H}{e^{\frac{i}{t}}-(1-H)} \mbox{。}
$$
The probability that the $\widetilde{m}_i$ different elements of the $i$-th column can NOT arrive at distinguished points after ~$\hat t-i$~ more iterations is $1-(1-1/t)^{\hat t-i}$. After discarding the chains not arriving at distinguished points, the number of different elements in the $i$-th column in a pre-computed matrix is
$$
{m}_i = \widetilde{m}_i \big(1-\big(1-\frac{1}{t}\big)^{\hat t-i}\big)\approx \widetilde{m}_i \cdot (1-e^{i/t-c}) \mbox.
$$
This completes the proof.
\qed

\begin{lemma}
During the online phase, $l$ pre-computed tables are processed in parallel. The probability of failure after the first $k$ iterating search, i.e. the probability of the $k+1$-th iterating search to be executed, is
$$
p_k = \mbox{exp}\bigg({~-\frac{\widetilde{m}_0tlH}{N}\cdot\int_{c-k/t}^{c}{~\frac{1-e^{u-c}}{e^u-(1-H)}}~du}\bigg).
$$
And the success probability is
$$
p = 1-p_{\hat{t}} = 1-\mbox{exp}\bigg({~-\frac{\widetilde{m}_0tlH}{N}\cdot\int_{0}^{c}{~\frac{1-e^{u-c}}{e^u-(1-H)}}~du}\bigg).
$$
\end{lemma}
\proof
For a random function $f$, we treat the columns of the pre-computed matrix as independent, i.e. all the different elements in every column are independent. Then
\begin{equation}
\label{eq:dp_rainbow_pk}
\begin{aligned}
p_k & = \prod_{i=1}^{k-1}(1-\frac {m_{\hat t-i}}N)^l \approx \prod_{i=1}^{k-1} e^{-\frac {m_{\hat t-i}l}N} & = \mbox{exp}\big({-\frac {l \sum_{i=1}^{k-1}m_{\hat t -i}}N} \big).
\end{aligned}
\end{equation}
And
$$
\begin{aligned}
& \sum_{i=1}^{k-1}m_{\hat t -i} = \sum_{i=\hat t-k+1}^{\hat t-1}m_{i}
=\sum_{i=\hat t-k+1}^{\hat t-1}{\frac{\widetilde{m}_0H}{e^{\frac{i}{t}}-(1-H)}\cdot(1-e^{i/t-c})} \\
&\approx \widetilde{m}_0tH~\int_{c-k/t}^{c}{~\frac{1-e^{u-c}}{e^u-(1-H)}}~du.
\end{aligned}
$$
Substitute it into Eq.~\ref{eq:dp_rainbow_pk} and we reach the claim after some simplification.

The probability of failure is the probability of NOT finding a pre-image after $\hat{t}$ iterating search. Thus the success probability is
$$
1-p_{\hat{t}} = 1-\mbox{exp}\bigg({~-\frac{\widetilde{m}_0tlH}{N}\cdot\int_{0}^{c}{~\frac{1-e^{u-c}}{e^u-(1-H)}}~du}\bigg).
$$
This completes the proof.
\qed
Denote by $E_{fa}(i)$ the expected number of false alarms during the $i$-th iterating search. Note that when online chain merges with multiple pre-computed chains, all these pre-computed chains need to be regenerated. Thus all pre-computed chains can be treated as independent.

It is expected that there are $\widetilde{m}_0(1-\frac{1}{t})^{j-1}\frac{1}{t}$ chains of length $j$ in every pre-computed table, and the probability that a pre-computed chain of length $j$ merges with an online chain of length $i$ is
$$
1-(1-\frac 1N)^{j -(\hat t-i)} \approx \frac{j-(\hat{t}-i)}{N},
$$
So,
$$
\begin{aligned}
E_{fa}(i) &= \sum_{j=\hat t- i+1}^{\hat t}  \widetilde{m}_0(1-\frac{1}{t})^{j-1}\frac{1}{t} \cdot \frac{j-(\hat t-i)}N \\
&=\frac{\widetilde{m}_0t}N[(1-\frac 1t)^{\hat t -i+2} \cdot(1-(1-\frac 1t)^{i-1})-\frac it(1-\frac 1t)^{\hat t +1})] ,
\end{aligned}
$$
which simplifies to
$$
\begin{aligned}
E_{fa}(i) \approx \frac{\widetilde{m}_0te^{-c}}{N}\cdot(e^{i/t}-\frac{i}{t}-1).
\end{aligned}
$$
To deal with every false alarms during the $i$-th iterating search, $(\hat t-i+1)$ function invocations are needed. Then the online time (denoted by number of function invocations)
\begin{equation}
\begin{aligned}
&T=\sum_{i=1}^{\hat t}{l[(i-1)+(\hat t-i+1) \cdot E_{fa}(i)]}\cdot p_{i-1} \\
&\approx t^2\int_{0}^{c}{\big[lv+\frac{m_0tl(c-v)e^{-c}}{N}(e^v-1-v)\big]\mbox{exp}\bigg({-\frac{\widetilde{m}_0tlH}{N}\cdot\int_{c-v}^{c}{\frac{1-e^{u-c}}{e^u-(1-H)}}}du\bigg)dv} \\
&=t^2\int_{0}^{c}{\big[lv+D_{pc}e^{-c}(c-v)(e^v-1-v)\big]\mbox{exp}\bigg({-\frac{D_{pc}H}{1-e^{-c}}\cdot\int_{c-v}^{c}{\frac{1-e^{u-c}}{e^u-(1-H)}}}du\bigg)dv}.
\end{aligned}
\end{equation}
The memory needed to store pre-computed tables is $M = lm_0$. Then the tradeoff coefficient of rainbow distinguished point method is
\begin{equation}
\begin{aligned}
&D_{tcr} = TM^2/N^2\\ &=D^2_{pc}\int_{0}^{c}{\big[lv+D_{pc}e^{-c}(c-v)(e^v-1-v)\big]\mbox{exp}\bigg({-\frac{D_{pc}H}{1-e^{-c}}\cdot\int_{c-v}^{c}{\frac{1-e^{u-c}}{e^u-(1-H)}}}du\bigg)dv},
\end{aligned}
\end{equation}
where $D_{pc} = m_0tl/N$ is the pre-computation coefficient of rainbow distinguished point method.

\subsection{Parameter Optimization}
We treat the number of pre-computation tables as discrete numbers. Using numerical method we compute the optimal parameter set, which minimize the tradeoff coefficient (as shown in Table \ref{table:vdprainbow}), when the expected pre-computation coefficient and success probability are specified.

For example, if a pre-computation time of $3.5N$ function invocations is expected, to achieve the success probability of $75\%$, it is best to pre-compute 2 rainbow distinguished point table with maximum chain length of $1.33t$. The expected tradeoff coefficient is $14.6280$.

The italic numbers in the table give optimal parameters, which does NOT achieve better tradeoff efficiency with more pre-computed time, due to the fact that the improvement of online time does not neutralize the efficiency reduction of more memory cost.
\begin{table}[h!]
\begin{center}
%\small{
\caption{Success probability, maximum chain length, pre-computation coefficient and tradeoff coefficient of rainbow distinguished point method}
\label{table:2}
\begin{tabular}{|c|c|c|c|c||c|c|c|c|c|}
\hline
l & p & $D_{pc}$ & c & $D_{tcr}$ & l & p & $D_{pc}$ & c & $D_{tcr}$  \\ \hline
1 & 0.6008  &   2.5   &  1.35  & 4.6907  & 1 & 0.5997  &   3   &  1.12  & 4.6395    \\ \hline
1 & 0.6506  &   2.5   &   1.66 & 7.0548  & 1 & 0.6505  &   3   &  1.36  & 6.8033  \\ \hline
1 & 0.6997  &   2.5  &  2.1  & 11.2609   & 1 & 0.7004  &   3   &  1.68  & 10.3554  \\ \hline
1 & 0.7507  &   2.5   &   2.9 &  21.4945   & 2 & 0.7504  &   3   &   1.59 &  15.4773  \\ \hline
2 & 0.8003  &   2.5   &   2.75 &  32.2432   & 2 & 0.8006  &   3   &   2.04 &  24.9292  \\ \hline
\hline
$\textit{1}$ & $\textit{0.6507}$  &   $\textit{3.5}$   &   $\textit{1.17}$ & $\textit{6.8485}$  & $\textit{1}$ & $\textit{0.7004}$  &   $\textit{4}$   & $\textit{1.25}$   & $\textit{10.1937}$       \\ \hline
1 & 0.6997  &   3.5   &  1.42  & 10.0658 & $\textit{1}$ & $\textit{0.7511}$  &  $\textit{4}$   & $\textit{1.55} $  &  $\textit{15.7085}$       \\ \hline
2 & 0.7506  &   3.5   &   1.33 &  14.6280 & 2 & 0.8002  &   4   &  2  &   21.0761     \\ \hline
2 & 0.8002  &   3.5   &  1.66  & 22.2033 & 2 & 0.8502  &   4   & 1.84   & 34.5087        \\ \hline
2 & 0.8500  &   3.5   &   2.21 &  38.5838 & 3 & 0.9006  &   4   &  2.23  &  66.9477      \\ \hline
\hline
$\textit{2}$ & $\textit{0.7016}$  & $\textit{4.5}$ & $\textit{0.86}$& $\textit{10.3470}$  & $\textit{2}$ & $\textit{0.7516}$ & $\textit{5}$ & $\textit{0.93}$ & $\textit{14.4699}$ \\ \hline
2 & 0.7520  &   4.5   &  1.03  & 14.3916 & 2 & 0.7993  &  5   & 1.12   &  20.3680       \\ \hline
2 & 0.8001  &   4.5   &   1.25 &  20.5882 & 2 & 0.8506  &   5   &  1.42  &   31.7488     \\ \hline
2 & 0.8509  &   4.5   &  1.60  & 32.7822 & 3 & 0.9005  &   5   & 1.64   & 54.9546        \\ \hline
3 & 0.8999  &   4.5   &   1.87 &  58.5550 & 4 & 0.9499 &   5   &  2.34  &  133.1831      \\ \hline
\hline
$\textit{2}$ & $\textit{0.7989}$ & $\textit{5.5}$ & $\textit{1.02}$ & $\textit{20.4230}$ & $\textit{2}$ & $\textit{0.8508}$ & $\textit{6}$ & $\textit{1.18}$ & $\textit{31.4500}$\\ \hline
2 & 0.8498  &   5.5   &   1.28 & 31.1537  & 3 & 0.9006  &   6   &  1.33  & 51.3959  \\ \hline
3 & 0.9000  &   5.5  &  1.46  & 52.3366   & 4 & 0.9497  &   6   &  1.79  & 108.2289  \\ \hline
4 & 0.9498  &   5.5   &   2.02 &  117.5015   & - & -  &   -   &  - &  -  \\ \hline
\hline
4 & 0.9799  &   10   &   1.48 & 183.2010 &  6 & 0.9901  &   10   &   1.64 & 281.7218  \\ \hline
7 & 0.9899  &   15   &   1.00 & 254.6985 &  - & -  &   -  &  - & - \\ \hline
\end{tabular}
%}
\label{table:vdprainbow}
\end{center}
\end{table}

\subsection{Experiment}
We use a truncated version of MD5 to verify our results. We set the search space $N=2^{24}$, number of chains $\widetilde{m}_0=262144$, chain length $t=512$, number of tables $l=1$ and maximum chain length $\hat{t}=1.8t$.

During the pre-computation phase the number of function invocations is $6.69N$ (theoretically $6.68N$). The pre-computed table contains $219083$ (theoretically $218812$) pre-computed chains. During the onlie phase, we tested 3000 random images. The average number of function invocations is $399804$ (theoretically $394023$). The success probability is approximately $87.6\%$ (theoretically $87.4\%$). The average number of false alarms is $511$ (theoretically $505$).

\subsection{Comparison}
Fixing a same pre-computed time and success probability, we compare the optimized tradeoff efficiency of rainbow distinguished point method with Hellman method, (Hellman) distinguished point method and rainbow method. See Table~\ref{table:vdprainbowcomp}. We can see that rainbow distinguished point method does NOT bring us better tradeoff efficiency.

\begin{table}[h!]
\begin{center}
\small{
\caption{Comparison of rainbow distinguished point method with other methods}
\label{table:1}
\begin{tabular}{|p{4cm}<{\centering}|p{2.2cm}<{\centering}|p{2.5cm}<{\centering}|c|}
\hline
Time memory tradeoff algorithms & Success probability & pre-computation coefficient & tradeoff curve\\ \hline
Hellman method & $P = 80\%$ & 2.1733 &$TM^2 = 3.11N^2$\\ \hline
Rainbow method &  $P = 80\%$ & 1.9814 &$TM^2 = 2.20N^2$\\ \hline
Hellman distinguished point method &  $P = 80\%$ & 2.9532 &$TM^2 = 11.58N^2$\\ \hline
Rainbow distinguished point method &  $P = 80\%$ & 3 &$TM^2 = 24.93N^2$\\ \hline
Hellman method &  $P = 90\%$ & 3.1093 &$TM^2 = 7.17N^2$\\ \hline
Rainbow method &  $P = 90\%$ & 2.8068 &$TM^2 = 4.68N^2$\\ \hline
Hellman distinguished point method &  $P = 90\%$ &  4.2250&$TM^2 = 26.59N^2$\\ \hline
Rainbow distinguished point method &  $P = 90\%$ &  4 &$TM^2 = 66.95N^2$\\ \hline
\end{tabular}
\label{table:vdprainbowcomp}
}
\end{center}
\end{table}

\bibliographystyle{plain}
\bibliography{reference}

\end{document}